\documentclass[]{emulateapj}

\bibpunct[, ]{(}{)}{;}{a}{}{,}
\shorttitle{Spectroscopy of SN\,2008D}
\shortauthors{Malesani et al.}

\begin{document}

\title{Early Spectroscopic Identification of SN\,2008D\altaffilmark{1}}
\altaffiltext{1}{Partly based on observations made with ESO telescopes
                 at the La Silla Paranal Observatory under program
                 080.D-0526, with the Nordic Optical Telescope, operated
		 on the island of La Palma jointly by Denmark, Finland,
		 Iceland, Norway, and Sweden, and with the United Kingdom
		 Infrared Telescope, which is operated by the Joint Astronomy
		 Centre on behalf of the Science and Technology Council of the
		 UK.}
\author{
D. Malesani\altaffilmark{2},
J. P. U. Fynbo\altaffilmark{2},
J. Hjorth\altaffilmark{2},
G. Leloudas\altaffilmark{2},
J. Sollerman\altaffilmark{2,3},
M. D. Stritzinger\altaffilmark{4,2},
P. M. Vreeswijk\altaffilmark{2},
D. J. Watson\altaffilmark{2},
J. Gorosabel\altaffilmark{5},
M. J. Micha\l{},owski\altaffilmark{2},
C. C. Th\"one\altaffilmark{2},
T. Augusteijn\altaffilmark{6},
D. Bersier\altaffilmark{7},
P. Jakobsson\altaffilmark{8},
A. O. Jaunsen\altaffilmark{9},
C. Ledoux\altaffilmark{10},
A. J. Levan\altaffilmark{11},
B. Milvang-Jensen\altaffilmark{2},
E. Rol\altaffilmark{12},
N. R. Tanvir\altaffilmark{12},
K. Wiersema\altaffilmark{12},
D. Xu\altaffilmark{2},
L. Albert\altaffilmark{13},
M. Bayliss\altaffilmark{14,15},
C. Gall\altaffilmark{2},
L. F. Grove\altaffilmark{2},
B. P. Koester\altaffilmark{14,15},
E. Leitet\altaffilmark{16},
T. Pursimo\altaffilmark{6},
I. Skillen\altaffilmark{17}
}
\altaffiltext{2}{Dark Cosmology Centre, Niels Bohr Institute, University of Copenhagen, Juliane Maries vej 30, 2100 Copenhagen \O, Denmark}
\altaffiltext{3}{Department of Astronomy, Stockholm University, 10691 Stockholm, Sweden}
\altaffiltext{4}{Las Campanas Observatory, Carnegie Institute of Science, Colina el Pino Casilla 601, La Serena, Chile}
\altaffiltext{5}{Instituto de Astrof\'\i{}sica de Andaluc\'\i{}a (IAA-CSIC), Apartado 3004, 18080 Granada, Spain}
\altaffiltext{6}{Nordic Optical Telescope, Apartado 474, 38700 Santa Cruz de La Palma, Spain}
\altaffiltext{7}{Astrophysics Research Institute, Liverpool John Moores University, Twelve Quays House, Egerton Wharf, Birkenhead, CH41 1LD, UK}
\altaffiltext{8}{Centre for Astrophysics and Cosmology, Science Institute, University of Iceland, Dunhagi 5, 107 Reykjav\'ik, Iceland}
\altaffiltext{9}{Institute of Theoretical Astrophysics, University of Oslo, P.O. Box 1029, Blindern, 0315 Oslo, Norway}
\altaffiltext{10}{European Southern Observatory, Avenida Alonso de Cordova 3107, Casilla 19001, Vitacura, Santiago, Chile}
\altaffiltext{11}{Department of Physics, University of Warwick, Coventry CV4 7AL, UK}
\altaffiltext{12}{Department of Physics and Astronomy, University of Leicester, Leicester, LE1 7RH, UK}
\altaffiltext{13}{Canada-France-Hawaii Telescope Corporation, 65-1238 Mamalahoa Highway, Kamuela, HI 96743, USA}
\altaffiltext{14}{Department of Astronomy and Astrophysics, University of Chicago, 5640 South Ellis Avenue, Chicago, IL 60637, USA}
\altaffiltext{15}{Kavli Institute for Cosmological Physics, 5640 South Ellis Avenue, Chicago, IL 60637, USA}
\altaffiltext{16}{Department of Physics and Astronomy, Uppsala University, Box 515, SE-751 20 Uppsala, Sweden}
\altaffiltext{17}{Isaac Newton Group, Apartado 321, 38700 Santa Cruz de La Palma, Canary Islands, Spain}

\begin{abstract}
\mbox{SN\,2008D} was discovered while following up an unusually bright X-ray
transient (XT) in the nearby spiral galaxy NGC\,2770. We present early optical
spectra (obtained 1.75 days after the XT) which allowed the first
identification of the object as a supernova (SN) at redshift $z = 0.007$. These
spectra were acquired during the initial declining phase of the light curve,
likely produced in the stellar envelope cooling after shock breakout, and
rarely observed. They exhibit a rather flat spectral energy distribution with
broad undulations, and a strong, W-shaped feature with minima at 3980 and 4190
\AA{} (rest frame). We also present extensive spectroscopy and photometry of
the SN during the subsequent photospheric phase. Unlike SNe associated with
gamma-ray bursts, \mbox{SN\,2008D} displayed prominent He features and is
therefore of Type Ib.
\end{abstract}

\keywords{supernovae: individual (SN 2008D)}

\section{Observations of SN\,2008D}\label{sec:intro}

On 2008 January 9.56 UT, while observing the supernova (SN) 2007uy in the
nearby spiral galaxy NGC\,2770 ($z = 0.007$), the X-Ray Telescope onboard
\textit{Swift} detected a bright X-ray transient (XT), with a peak luminosity
of $6 \times 10^{43}$~erg~s$^{-1}$ and a duration of about 10~minutes
\citep{Soderberg08a}. Its power-law spectrum and light curve shape were
reminiscent of gamma-ray bursts (GRBs) and X-ray flashes, but the energy
release was at least 2 orders of magnitude lower than for typical and even
subluminous GRBs, also allowing for beaming (e.g.,
\citealt{Amati06,Ghirlanda07}). The discovery of the XT prompted the search
for, and discovery of, an optical counterpart \citep{Deng08,Valenti08}.

We performed spectroscopy of the source as soon as possible, starting 1.75 days
after the XT, using the FORS2 spectrograph on the ESO Very Large Telescope
(VLT). Subsequent spectroscopic monitoring of the object was carried out at the
Nordic Optical Telescope (NOT) and the William Herschel Telescope (WHT). All
spectra have been reduced using standard techniques. On January 18.22 UT (8.65
days after the XT) we secured a high-resolution spectrum using the UVES
instrument on the VLT. For this observation, we adopted the ESO CPL pipeline
(v3.3.1), and flux calibration was performed using the master response curves.
The observing log of the spectra is reported in Table~\ref{speclog}.

Imaging observations were conducted using the NOT, the VLT, the Liverpool 
Telescope (LT) and the United Kingdom Infrared Telescope (UKIRT). Image 
reduction was carried out using standard techniques. For photometric 
calibration, we observed optical standard star fields on five different nights,
and defined a local sequence in the NGC\,2770 field. In the near-infrared we
used Two Micron All Sky Survey stars as calibrators. Magnitudes were computed
using small apertures, and subtracting the background as measured in an annulus
around the SN position. The contribution from the underlying host galaxy light
was always negligible, as also apparent from archival Sloan Digital Sky Survey
images. Our photometric results are listed in Table~\ref{tab:photo}.

One of our spectra covers the nucleus of NGC\,2770, allowing a precise
measurement of its redshift: $z = 0.0070 \pm 0.0009$. This is slightly larger
than the value listed in the NASA Extragalactic Database ($z = 0.0065$). For
$H_0 = 71$ km~s$^{-1}$~Mpc$^{-1}$, the luminosity distance is 29.9 Mpc.

\begin{figure}
\includegraphics[width=\columnwidth]{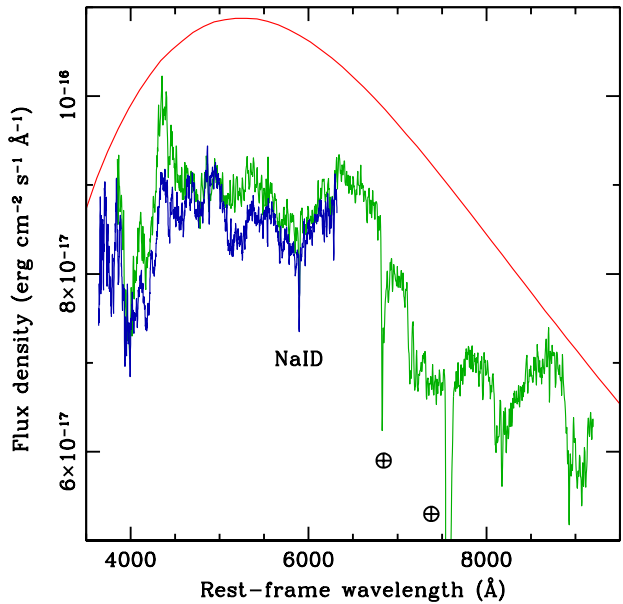}
\caption{Our two earliest spectra obtained 1.75 days after the XT, when the
cooling envelope emission was dominating the observed light. The spectra cover
the wavelength ranges 3600--6300~\AA{} (grism 600B, blue curve) and
3800--9200~\AA{} (grism 300V, green curve). For comparison, the red line shows
a blackbody spectrum with temperature $T = 15,000$~K, reddened assuming $E(B-V)
= 0.8$~mag (Section~\ref{sec:AV}). The \ion{Na}{1}~D narrow absorption from the
interstellar medium in NGC\,2770 is also noted, as well as the two strong
telluric features (marked with `$\oplus$').\label{fg:earlyspec}}
\end{figure}

\section{Results}\label{sec:results}

\subsection{The Early Spectrum of SN\,2008D}\label{sec:earlyspec}

Our first optical spectrum of the transient source (Figure~\ref{fg:earlyspec})
exhibits \ion{Na}{1}~D absorption lines at $z = 0.0070$, thus establishing its
extragalactic nature. Broad features are also apparent across the whole
spectrum ($\mbox{FWHM} = (1\mbox{--}3) \times 10^4$~km~s$^{-1}$), which led us
to identify the object as a core-collapse SN \citep{Malesani08a}.
\citet{Soderberg08a,Soderberg08b} describe nearly simultaneous spectra as
featureless, probably due to their smaller covered wavelength range
($4500\mbox{--}8000$~\AA). \citet{Modjaz08a} report features consistent with
those in our data.

We initially classified the SN as a very young Type Ib/c, based on the absence
of conspicuous Si and H lines \citep{Malesani08a}. As the spectrum  is among
the earliest observed for any SN, comparable only to the very first spectrum of
\mbox{SN\,1987A} \citep{Menzies87}, there is no obvious resemblance with known
SN spectra. It is notable, however, that the earliest spectrum of the Type-Ic
\mbox{SN\,1994I} was essentially flat with broad, low-amplitude undulations
(though the covered wavelength range was limited; \citealt{Filippenko95}).
Early spectra, also mostly featureless, are available for the H-rich Type-IIP
\mbox{SN\,2006bp} \citep{Quimby06}. \citet{Dessart08} interpret them in terms
of high temperature and ionization.

A striking feature in the spectrum is a conspicuous W-shaped absorption with
minima at 3980 and 4190 \AA{} (rest frame). It was detected using two different
instrument setups (Figure~\ref{fg:earlyspec}), and also reported by
\citet{Modjaz08a}. If interpreted as due to P~Cyg profiles, the inferred
expansion velocity is $\sim 15,000$ km~s$^{-1}$, computed from the position of
the bluest part compared to the peak. Its origin is unclear, although,
following \citet{Quimby07}, \citet{Modjaz08a} propose that it is due to a
combination of \ion{C}{3}, \ion{N}{3}, and \ion{O}{3}. Interpreting the broad
absorption at $\sim 5900$~\AA{} as \ion{Si}{2} $\lambda\lambda$ 6347, 6371,
some ejecta reached $\sim 22,000$ km~s$^{-1}$. Such large velocities have been
seen only in broad-lined (BL) Type-Ic SNe, at significantly later stages
\citep{Patat01,Hjorth03,Mazzali06}.

\begin{figure}
\includegraphics[width=\columnwidth]{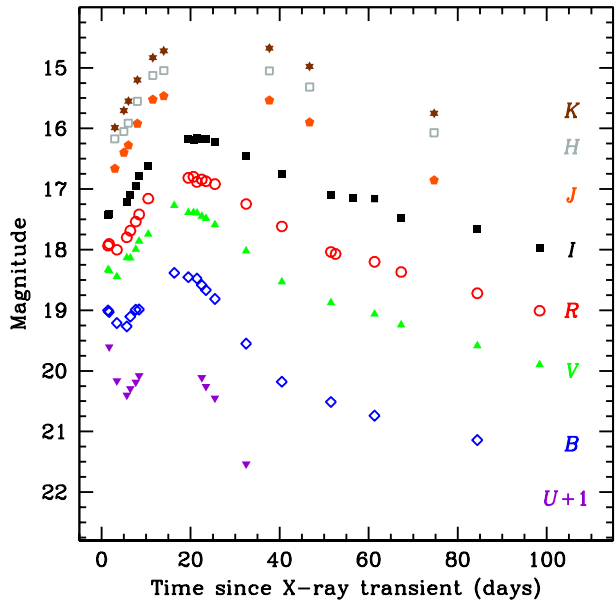}
\caption{Optical and near-infrared light curves of \mbox{SN\,2008D}. The data
points are not corrected for extinction. Error bars are smaller than symbols
and have been omitted.\label{fg:lc}}
\end{figure}

\subsection{The Photospheric Phase}

\begin{figure}
\centering\includegraphics[width=\columnwidth]{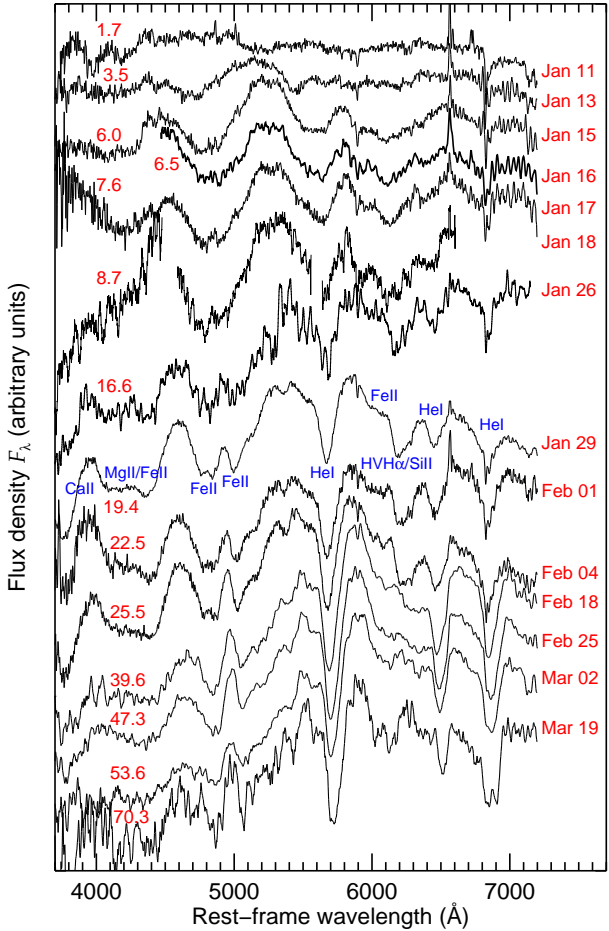}
\caption{
Spectral evolution of \mbox{SN\,2008D} from 1.75 days to 7 weeks after
explosion. On the left, the time in days since the XT is noted. The spectrum
marked as ``Jan 16'' is the average of those taken on January 15.95 and 16.26
UT. Close to the January 29 track we have indicated the most likely
identification of the main features (HVH$\alpha$ stands for ``high-velocity
H$\alpha$''). The narrow emission line at 6560~\AA{} is residual H$\alpha$ from
the SN host galaxy. The \ion{He}{1} feature around 6800~\AA{} is affected by
the B-band atmospheric absorption.\label{fg:specevo}}
\end{figure}

Figure~\ref{fg:lc} shows the optical and near-infrared light curves of
\mbox{SN\,2008D}. In the first days after the XT, the flux dropped faster in
the bluer bands, with the color becoming progressively redder. This can be
interpreted as due to the stellar envelope cooling after the shock breakout
\citep{Soderberg08a}. Our first spectra were taken during this stage, before
energy deposition by radioactive nuclei became dominant, hence the physical
conditions of the emitting material might be different than later. We note that
the W-shaped absorption discussed in Section~\ref{sec:earlyspec} was no longer
visible from 3.5~days after the XT onward (Fig.~\ref{fg:specevo}).

\begin{figure}
\includegraphics[width=\columnwidth]{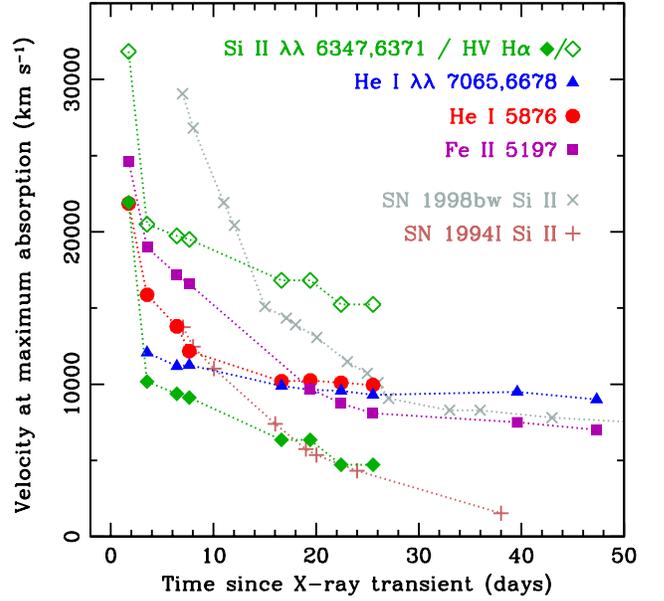}
\caption{
Velocity at maximum absorption of several transitions computed with SYNOW.
Diamonds indicate the velocity for the 6200~\AA{} line if interpreted as
\ion{Si}{2} (filled symbols) or HV H$\alpha$ (open symbols). The latter
interpretation is unlikely due to the lack of corresponding H$\beta$ (see also
\citealt{Tanaka08}).
\mbox{SN\,2008D} shows velocities lower than the prototypical hypernova
\mbox{SN\,1998bw} \citep{Patat01} and comparable to \mbox{SN\,1994I}
\citep{Sauer06}.\label{fg:photov}}
\end{figure}

The later spectra, acquired during the radioactivity-powered phase and
extending over more than two months in time, established \mbox{SN\,2008D} as a
Type-Ib SN \citep{Modjaz08b}. From January 17 and onward unambiguous He lines
are observed (Figure~\ref{fg:specevo}), consistent with other reports
\citep{Modjaz08a,Valenti08b,Soderberg08a,Tanaka08,Mazzali08}. In
Figure~\ref{fg:photov}, we plot the velocities at maximum absorption of a few
transitions determined using the SYNOW code \citep{Fisher99}. For comparison,
we also plot the \ion{Si}{2} velocity of the BL \mbox{SN\,1998bw}
\citep{Patat01} and of the normal Type-Ic \mbox{SN\,1994I} \citep{Sauer06},
showing that the velocities of \mbox{SN\,2008D} are lower than those of BL SNe.

\subsection{Dust Extinction}\label{sec:AV}

It follows from the detection of strong \ion{Na}{1}~D with an equivalent width
(EW) of 1.3~\AA{} that the extinction toward \mbox{SN\,2008D} is substantial
in NGC\,2770. Our best estimate of the reddening comes from comparing the
colors of \mbox{SN\,2008D} with those of stripped-envelope SNe, which have $V-R
\approx R-I \approx 0.1$ around maximum (e.g.,
\citealt{Folatelli06,Richmond96,Galama98}). The resulting reddening is $E(B-V)
= 0.8$~mag, corresponding to an extinction $A_V = 2.5$~mag (using the
extinction law by \citealt{Cardelli89} with $R_V = 3.1$).

A large dust content is supported by absorption features in our high-resolution
spectrum (see also \citealt{Soderberg08a}). The \ion{Na}{1}~D1 absorption line
indicates a multicomponent system, spanning a velocity range of 43 km~s$^{-1}$,
which sets a lower limit $E(B-V) > 0.2$ mag \citep[their Figure~4]{Munari97}.
The \ion{Na}{1}~D versus $E(B-V)$ relation for SNe \citep{Turatto03} suggests
$0.2~\mbox{mag} \la E(B-V) \la 0.6$~mag. Diffuse interstellar bands (DIBs) are
also detected at 5781.2, 5797.8, and 6283.9 \AA{} (rest frame). Their EWs
suggest $0.5~\mbox{mag} \la E(B-V) \la 2$~mag \citep{Cox05}. A dusty
environment has been directly revealed through millimeter imaging of NGC\,2770
\citep{Gorosabel08}. Last, the hydrogen column density in the X-ray spectrum of
the XT is $N_{\rm H} = 6.9^{+1.8}_{-1.5} \times10^{21}$ cm$^{-2}$ (assuming
Solar abundances; \citealt{Soderberg08a}). The gas-to-dust ratio is
$N_\mathrm{H}/A_V = 2.8\times10^{21}$ cm$^{-2}$ mag$^{-1}$, close to the
Galactic value $1.7\times 10^{21}$ cm$^{-2}$~mag$^{-1}$ \citep{Predehl95}.

\section{Discussion}\label{sec:discussion}

The precise explosion epoch is so far only known for a few Type-II SNe, thanks
to either the detection of the neutrino signal (\mbox{SN\,1987A};
\citealt{Hirata87,Bionta87}) or of the UV flash by \textit{Galaxy Evolution
Explorer} \citep{Schawinski08,Gezari08}, and for BL Type-Ic SNe associated with
GRBs (e.g., \citealt{Galama98,Hjorth03,Stanek03,Campana06}). \mbox{SN\,2008D}
is the first Type-Ib SN with a precisely constrained explosion epoch, since the
XT is expected to occur less than 1~hr after the stellar collapse
\citep{Li07,Waxman07}. The nature of the XT --- shock breakout versus
relativistic ejecta --- is still debated
\citep{Soderberg08a,Xu08,Li08,Mazzali08,ChFr08}, thus it is unclear whether
this phenomenon is common.

The light-curve evolution of \mbox{SN\,2008D} is very similar to that of
\mbox{SN\,1999ex} \citep{Stritzinger02}. The initial fading can be interpreted
as due to the envelope cooling through expansion following the initial X-ray/UV
flash \citep{Stritzinger02,Campana06,Soderberg08a}. The subsequent
rebrightening is due to the energy released by radioactive material in the
inner layers and gradually reaching the optically thin photosphere.

\begin{figure}
\includegraphics[width=\columnwidth]{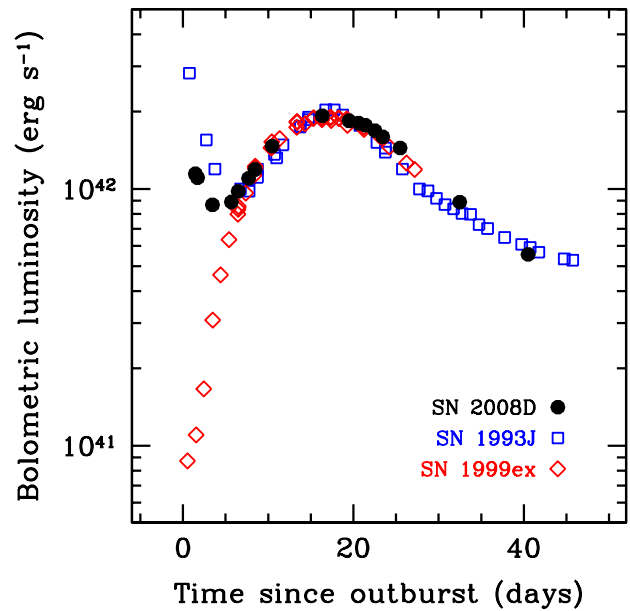}
\caption{Bolometric light curves of \mbox{SN\,2008D} (Type Ib),
\mbox{SN\,1993J} (Type IIb), and \mbox{SN\,1999ex} (Type Ib). Extinctions
corresponding to $E(B-V) = 0.8$, 0.19 and 0.30 mag were assumed. The
cooling envelope phase was visible only in the $U$ band for
\mbox{SN\,1999ex}.\label{fg:lcbolo}}
\end{figure}

From our $UBVRI$ data, we constructed the bolometric light curve of
\mbox{SN\,2008D} (Figure~\ref{fg:lcbolo}). For comparison we also show the two
other He-rich SNe caught during the early cooling phase: \mbox{SN\,1993J}
\citep{Richmond94} and \mbox{SN\,1999ex} \citep{Stritzinger02}. Strikingly, the
three SNe had very similar light curves during the photospheric phase. Given
its peak luminosity, \mbox{SN\,2008D} synthesized about $0.09 M_\odot$ of
$^{56}$Ni based on Arnett's rule \citep{Arnett82}. The emission during the
early cooling phase, however, varied substantially for the three SNe (seen only
in $U$ for \mbox{SN\,1999ex}). Furthermore, significant radiation may be
emitted blueward of the $U$ band during this early phase, so that the
bolometric values may be underestimated.

\mbox{SN\,2008D} has different properties from the SNe associated with GRBs,
namely the presence of He in the ejecta, a lower peak luminosity ($\approx
1.5$~mag), lower expansion velocities (a factor of $\approx 2$), and lower
$^{56}$Ni mass (a factor of $\approx 5$). The SN environment is also unlike
that of GRBs \citep{Soderberg08a,Thoene08}. Whether these two kinds of
high-energy transients are separate phenomena or form a continuum is unclear.
Addressing this issue will require theoretical modeling and an enlarged sample.
The discovery of a short-lived XT associated with an ordinary Type-Ib SN opens
the possibility of accessing the very early phases of ordinary SNe, which will
provide new insights into SN physics. Future X-ray sky-scanning experiments,
such as Lobster or eROSITA, may turn out, rather unexpectedly, ideally suited
to examine this issue, alerting us to the onset of many core collapse SNe.

\acknowledgments The Dark Cosmology Centre is supported by the DNRF. J.G. is
supported by the Spanish research programs AYA2004-01515 and
ESP2005-07714-C03-03 and P.M.V. by the EU under a Marie Curie Intra-European
Fellowship, contract MEIF-CT-2006-041363. P.J. acknowledges support by a Marie
Curie European Re-integration Grant within the 7th European Community Framework
Program under contract number PERG03-GA-2008-226653, and a Grant of Excellence
from the Icelandic Research Fund. We thank M.{} Modjaz and M.{} Tanaka for
discussion, and the observers at VLT, NOT, WHT, UKIRT, in particular A.{}
Djupvik, S.{} Niemi, A.{} Somero, T.{} Stanke, J.{} Telting, H.{} Uthas, and
C.{} Villforth.

{}


\begin{deluxetable}{@{}lrll@{}}
\tablecaption{Log of spectroscopic observations. Phases are computed relative to the XT onset.\label{speclog}}
\tablehead{
Epoch (UT) &
Phase (days) &
Telescope/instrument &
Exposure time (s)
}
\startdata
Jan 11.31 &   1.75 & VLT/FORS2+G300V  & $1\times600$  \\
Jan 11.32 &   1.76 & VLT/FORS2+G600B  & $1\times900$  \\
Jan 13.07 &   3.51 & NOT/ALFOSC+G4    & $3\times1200$ \\
Jan 15.17 &   5.61 & NOT/ALFOSC+G4    & $3\times1200$ \\
Jan 15.95 &   6.39 & NOT/ALFOSC+G4    & $1\times1200$ \\
Jan 16.26 &   6.70 & NOT/ALFOSC+G4    & $1\times1200$ \\
Jan 17.20 &   7.64 & NOT/ALFOSC+G4    & $3\times1200$ \\
Jan 18.22 &   8.66 & VLT/UVES+Dic\#1  & $1\times3600$ \\
Jan 26.15 &  16.59 & WHT/ISIS+R300B/R316R& $6\times600$ \\
Jan 29.01 &  19.45 & NOT/ALFOSC+G4    & $3\times1200$ \\
Feb 01.02 &  53.46 & NOT/ALFOSC+G4    & $2\times900$  \\
Feb 04.05 &  56.49 & NOT/ALFOSC+G4    & $2\times900$  \\
Feb 18.15 &  70.59 & NOT/ALFOSC+G4    & $3\times1200$ \\
Feb 25.88 &  78.32 & NOT/ALFOSC+G4    & $2\times1500$ \\
Mar 02.18 & 112.62 & NOT/ALFOSC+G4    & $3\times1200$ \\
Mar 18.87 & 129.31 & NOT/ALFOSC+G4    & $3\times900$
\enddata
\end{deluxetable}

\begin{deluxetable}{@{}llccl@{}}
\tablecaption{Log of optical and near-infrared imaging observations. For the
$UBVRI$ data, magnitudes do not include the zeropoint calibration error of
0.10, 0.03, 0.04, 0.03, and 0.04 mag, respectively.\label{tab:photo}}
\tablehead{
Epoch (UT) &Phase (days) &Filter & Magnitude    &Instrument
}
\startdata
Jan 11.26528 & 01.70082 & $U$ & 18.60$\pm$0.02 & NOT+StanCam \\
Jan 13.01426 & 03.44980 & $U$ & 19.16$\pm$0.05 & NOT+ALFOSC  \\
Jan 15.22931 & 05.66485 & $U$ & 19.40$\pm$0.05 & NOT+ALFOSC  \\
Jan 16.03116 & 06.46670 & $U$ & 19.29$\pm$0.07 & NOT+ALFOSC  \\
Jan 17.25019 & 07.68573 & $U$ & 19.19$\pm$0.07 & NOT+ALFOSC  \\
Jan 18.00103 & 08.43657 & $U$ & 19.08$\pm$0.04 & NOT+StanCam \\
Feb 01.06841 & 22.50395 & $U$ & 19.11$\pm$0.04 & NOT+ALFOSC  \\
Feb 02.00892 & 23.44446 & $U$ & 19.26$\pm$0.05 & NOT+ALFOSC  \\
Feb 04.03992 & 25.47546 & $U$ & 19.45$\pm$0.04 & NOT+ALFOSC  \\
Feb 11.03650 & 32.47204 & $U$ & 20.54$\pm$0.08 & NOT+ALFOSC  \\ \hline
Jan 11.04261 & 01.47815 & $B$ & 19.00$\pm$0.01 & NOT+StanCam \\
Jan 11.26873 & 01.70427 & $B$ & 19.03$\pm$0.02 & NOT+StanCam \\
Jan 13.01980 & 03.45534 & $B$ & 19.21$\pm$0.02 & NOT+ALFOSC  \\
Jan 15.23748 & 05.67302 & $B$ & 19.27$\pm$0.04 & NOT+ALFOSC  \\
Jan 16.03701 & 06.47255 & $B$ & 19.11$\pm$0.01 & NOT+ALFOSC  \\
Jan 17.25617 & 07.69171 & $B$ & 18.99$\pm$0.04 & NOT+ALFOSC  \\
Jan 18.00862 & 08.44416 & $B$ & 18.99$\pm$0.05 & NOT+StanCam \\
Jan 25.90236 & 16.33790 & $B$ & 18.38$\pm$0.08 & LT+RATCam   \\
Jan 29.05104 & 19.48658 & $B$ & 18.45$\pm$0.01 & NOT+ALFOSC  \\
Jan 30.98094 & 21.41648 & $B$ & 18.48$\pm$0.07 & NOT+ALFOSC  \\
Feb 01.07844 & 22.51398 & $B$ & 18.59$\pm$0.01 & NOT+ALFOSC  \\
Feb 02.01195 & 23.44749 & $B$ & 18.67$\pm$0.01 & NOT+ALFOSC  \\
Feb 04.04304 & 25.47858 & $B$ & 18.81$\pm$0.03 & NOT+ALFOSC  \\
Feb 11.04021 & 32.47575 & $B$ & 19.55$\pm$0.03 & NOT+ALFOSC  \\
Feb 18.98034 & 40.41588 & $B$ & 20.18$\pm$0.09 & NOT+ALFOSC  \\
Mar 01.09311 & 51.52865 & $B$ & 20.51$\pm$0.03 & NOT+ALFOSC  \\
Mar 10.88157 & 61.31711 & $B$ & 20.74$\pm$0.04 & NOT+MOSCA   \\
Apr 02.93574 & 84.37128 & $B$ & 21.14$\pm$0.05 & NOT+MOSCA   \\ \hline
Jan 11.02120 & 01.45674 & $V$ & 18.33$\pm$0.01 & NOT+StanCam \\
Jan 11.27136 & 01.70690 & $V$ & 18.35$\pm$0.06 & NOT+StanCam \\
Jan 13.02744 & 03.46298 & $V$ & 18.45$\pm$0.01 & NOT+ALFOSC  \\
Jan 15.24242 & 05.67796 & $V$ & 18.13$\pm$0.05 & NOT+ALFOSC  \\
Jan 16.04183 & 06.47737 & $V$ & 18.14$\pm$0.01 & NOT+ALFOSC  \\
Jan 17.26096 & 07.69650 & $V$ & 17.99$\pm$0.02 & NOT+ALFOSC  \\
Jan 18.01366 & 08.44920 & $V$ & 17.86$\pm$0.13 & NOT+StanCam \\
Jan 20.02994 & 10.46548 & $V$ & 17.75$\pm$0.03 & NOT+StanCam \\
Jan 25.90873 & 16.34427 & $V$ & 17.27$\pm$0.10 & LT+RATCam   \\
Jan 29.05856 & 19.49410 & $V$ & 17.39$\pm$0.01 & NOT+ALFOSC  \\
Jan 30.19193 & 20.62747 & $V$ & 17.39$\pm$0.02 & NOT+ALFOSC  \\
Jan 30.97634 & 21.41188 & $V$ & 17.40$\pm$0.08 & NOT+ALFOSC  \\
Feb 01.08387 & 22.51941 & $V$ & 17.45$\pm$0.01 & NOT+ALFOSC  \\
Feb 02.01399 & 23.44953 & $V$ & 17.49$\pm$0.01 & NOT+ALFOSC  \\
Feb 04.04479 & 25.48033 & $V$ & 17.59$\pm$0.01 & NOT+ALFOSC  \\
Feb 11.04238 & 32.47792 & $V$ & 18.02$\pm$0.01 & NOT+ALFOSC  \\
Feb 18.98778 & 40.42332 & $V$ & 18.53$\pm$0.04 & NOT+ALFOSC  \\
Mar 01.09818 & 51.53372 & $V$ & 18.87$\pm$0.02 & NOT+ALFOSC  \\
Mar 10.88568 & 61.32122 & $V$ & 19.06$\pm$0.05 & NOT+MOSCA   \\
Mar 17.86086 & 68.29640 & $V$ & 19.24$\pm$0.08 & NOT+ALFOSC  \\
Apr 02.94851 & 84.38405 & $V$ & 19.58$\pm$0.03 & NOT+MOSCA   \\
Apr 16.95106 & 98.38660 & $V$ & 19.90$\pm$0.07 & NOT+StanCam
\enddata
\end{deluxetable}

\addtocounter{table}{-1}
\begin{deluxetable}{@{}llccl@{}}
\tablecaption{(continued)}
\tablehead{
Epoch (UT) &Phase (days) &Filter & Magnitude    &Instrument
}
\startdata
Jan 11.00922 & 01.44476 & $R$ & 17.94$\pm$0.01 & NOT+StanCam \\
Jan 11.26190 & 01.69744 & $R$ & 17.91$\pm$0.02 & NOT+StanCam \\
Jan 11.30417 & 01.73971 & $R$ & 17.94$\pm$0.02 & VLT+FORS2   \\
Jan 13.02367 & 03.45921 & $R$ & 18.00$\pm$0.01 & NOT+ALFOSC  \\
Jan 15.24730 & 05.68284 & $R$ & 17.80$\pm$0.01 & NOT+ALFOSC  \\
Jan 16.04659 & 06.48213 & $R$ & 17.69$\pm$0.01 & NOT+ALFOSC  \\
Jan 17.26593 & 07.70147 & $R$ & 17.54$\pm$0.01 & NOT+ALFOSC  \\
Jan 18.01774 & 08.45328 & $R$ & 17.42$\pm$0.01 & NOT+StanCam \\
Jan 20.04708 & 10.48262 & $R$ & 17.16$\pm$0.04 & NOT+StanCam \\
Jan 29.06847 & 19.50401 & $R$ & 16.82$\pm$0.01 & NOT+ALFOSC  \\
Jan 30.21722 & 20.65276 & $R$ & 16.80$\pm$0.02 & NOT+ALFOSC  \\
Jan 30.97162 & 21.40716 & $R$ & 16.88$\pm$0.05 & NOT+ALFOSC  \\
Feb 01.08694 & 22.52248 & $R$ & 16.84$\pm$0.01 & NOT+ALFOSC  \\
Feb 02.01603 & 23.45157 & $R$ & 16.87$\pm$0.01 & NOT+ALFOSC  \\
Feb 04.04619 & 25.48173 & $R$ & 16.92$\pm$0.01 & NOT+ALFOSC  \\
Feb 11.04420 & 32.47974 & $R$ & 17.25$\pm$0.01 & NOT+ALFOSC  \\
Feb 18.99276 & 40.42830 & $R$ & 17.62$\pm$0.01 & NOT+ALFOSC  \\
Mar 01.10190 & 51.53744 & $R$ & 18.04$\pm$0.01 & NOT+ALFOSC  \\
Mar 02.12923 & 52.56477 & $R$ & 18.07$\pm$0.02 & NOT+ALFOSC  \\
Mar 10.88805 & 61.32359 & $R$ & 18.20$\pm$0.02 & NOT+MOSCA   \\
Mar 17.86648 & 68.30202 & $R$ & 18.37$\pm$0.05 & NOT+ALFOSC  \\
Apr 02.95326 & 84.38880 & $R$ & 18.72$\pm$0.04 & NOT+MOSCA   \\
Apr 16.94566 & 98.38120 & $R$ & 19.01$\pm$0.04 & NOT+StanCam \\ \hline
Jan 11.03211 & 01.46765 & $I$ & 17.43$\pm$0.01 & NOT+StanCam \\
Jan 11.27381 & 01.70935 & $I$ & 17.40$\pm$0.01 & NOT+StanCam \\
Jan 15.25279 & 05.68833 & $I$ & 17.21$\pm$0.01 & NOT+ALFOSC  \\
Jan 16.05135 & 06.48689 & $I$ & 17.09$\pm$0.01 & NOT+ALFOSC  \\
Jan 17.27084 & 07.70638 & $I$ & 16.95$\pm$0.01 & NOT+ALFOSC  \\
Jan 18.02196 & 08.45750 & $I$ & 16.79$\pm$0.01 & NOT+StanCam \\
Jan 20.05173 & 10.48727 & $I$ & 16.62$\pm$0.07 & NOT+StanCam \\
Jan 29.07600 & 19.51154 & $I$ & 16.18$\pm$0.01 & NOT+ALFOSC  \\
Jan 30.22517 & 20.66071 & $I$ & 16.19$\pm$0.05 & NOT+ALFOSC  \\
Jan 30.99911 & 21.43465 & $I$ & 16.16$\pm$0.04 & NOT+ALFOSC  \\
Feb 01.09031 & 22.52585 & $I$ & 16.17$\pm$0.01 & NOT+ALFOSC  \\
Feb 02.01851 & 23.45405 & $I$ & 16.18$\pm$0.01 & NOT+ALFOSC  \\
Feb 04.04764 & 25.48318 & $I$ & 16.23$\pm$0.01 & NOT+ALFOSC  \\
Feb 11.04589 & 32.48143 & $I$ & 16.45$\pm$0.01 & NOT+ALFOSC  \\
Feb 18.99647 & 40.43201 & $I$ & 16.76$\pm$0.01 & NOT+ALFOSC  \\
Mar 01.07881 & 51.51435 & $I$ & 17.09$\pm$0.01 & NOT+ALFOSC  \\
Mar 06.11299 & 56.54853 & $I$ & 17.15$\pm$0.03 & VLT+FORS1   \\
Mar 10.89042 & 61.32596 & $I$ & 17.16$\pm$0.04 & NOT+MOSCA   \\
Mar 17.87137 & 68.30691 & $I$ & 17.48$\pm$0.06 & NOT+ALFOSC  \\
Apr 02.95712 & 84.39266 & $I$ & 17.65$\pm$0.07 & NOT+MOSCA   \\
Apr 16.95661 & 98.39215 & $I$ & 17.97$\pm$0.02 & NOT+StanCam \\ \hline
Jan 12.555   & 02.99100 & $J$ & 16.66$\pm$0.02 & UKIRT+UFTI  \\
Jan 14.598   & 05.03400 & $J$ & 16.40$\pm$0.02 & UKIRT+UFTI  \\
Jan 15.457   & 05.89300 & $J$ & 16.28$\pm$0.02 & UKIRT+UFTI  \\
Jan 17.645   & 08.08100 & $J$ & 15.92$\pm$0.02 & UKIRT+UFTI  \\
Jan 21.004   & 11.44000 & $J$ & 15.52$\pm$0.02 & NOT+NOTCam  \\
Jan 23.576   & 14.01200 & $J$ & 15.47$\pm$0.02 & UKIRT+UFTI  \\
Feb 16.240   & 37.67600 & $J$ & 15.54$\pm$0.02 & UKIRT+WFCAM \\
Feb 25.295   & 46.73100 & $J$ & 15.90$\pm$0.02 & UKIRT+WFCAM \\
Mar 24.250   & 74.69000 & $J$ & 16.85$\pm$0.02 & UKIRT+WFCAM \\ \hline
Jan 12.555   & 02.99100 & $H$ & 16.17$\pm$0.02 & UKIRT+UFTI  \\
Jan 14.598   & 05.03400 & $H$ & 16.05$\pm$0.02 & UKIRT+UFTI  \\
Jan 15.457   & 05.89300 & $H$ & 15.91$\pm$0.02 & UKIRT+UFTI  \\
Jan 17.645   & 08.08100 & $H$ & 15.55$\pm$0.02 & UKIRT+UFTI  \\
Jan 21.004   & 11.44000 & $H$ & 15.13$\pm$0.02 & NOT+NOTCam  \\
Jan 23.576   & 14.01200 & $H$ & 15.05$\pm$0.02 & UKIRT+UFTI  \\
Feb 16.240   & 37.67600 & $H$ & 15.05$\pm$0.02 & UKIRT+WFCAM \\
Feb 25.295   & 46.73100 & $H$ & 15.32$\pm$0.02 & UKIRT+WFCAM \\
Mar 24.250   & 74.69000 & $H$ & 16.07$\pm$0.02 & UKIRT+WFCAM \\ \hline
Jan 12.555   & 02.99100 & $K$ & 15.99$\pm$0.02 & UKIRT+UFTI  \\
Jan 14.598   & 05.03400 & $K$ & 15.71$\pm$0.02 & UKIRT+UFTI  \\
Jan 15.457   & 05.89300 & $K$ & 15.55$\pm$0.02 & UKIRT+UFTI  \\
Jan 17.645   & 08.08100 & $K$ & 15.20$\pm$0.02 & UKIRT+UFTI  \\
Jan 21.004   & 11.44000 & $K$ & 14.83$\pm$0.02 & NOT+NOTCam  \\
Jan 23.576   & 14.01200 & $K$ & 14.72$\pm$0.02 & UKIRT+UFTI  \\
Feb 16.240   & 37.67600 & $K$ & 14.68$\pm$0.02 & UKIRT+WFCAM \\
Feb 25.295   & 46.73100 & $K$ & 14.98$\pm$0.02 & UKIRT+WFCAM \\
Mar 24.250   & 74.69000 & $K$ & 15.75$\pm$0.02 & UKIRT+WFCAM
\enddata
\end{deluxetable}


\begin{thebibliography}{}
\bibitem[Amati(2006)]{Amati06} Amati, L.\ 2006, \mnras, 372, 233
\bibitem[Arnett(1982)]{Arnett82} Arnett, W.~D. 1982, \apj, 253, 785
\bibitem[Bionta et al.(1987)]{Bionta87} Bionta, R.~M., et al.\ 1987, \prl, 58, 1494
\bibitem[Campana et al.(2006)]{Campana06} Campana, S., et al.\ 2006, \nat, 442, 1008
\bibitem[Cardelli et al.(1989)]{Cardelli89} Cardelli, J.~A., Clayton, G.~C., \& Mathis, J.~S.\ 1989, \apj, 345, 245
\bibitem[Chevalier \& Fransson(2008)]{ChFr08} Chevalier, R., \& Fransson, C.\ 2008, \apj, 683, L135
\bibitem[Cox et al.(2005)]{Cox05} Cox, N.~L.~J., Kaper,~L., Foing, B.~H., \& Ehrenfreund, P.\ 2005, \aap, 438, 187
\bibitem[Deng \& Zhu(2008)]{Deng08} Deng, J., \& Zhu, Y.\ 2008, GCN Circ. 7160
\bibitem[Dessart et al.(2008)]{Dessart08} Dessart, L., et al.\ 2008, \apj, 675, 644
\bibitem[Filippenko et al.(1995)]{Filippenko95} Filippenko, A.~V., et al.\ 1995, \apj, 450, L11
\bibitem[Fisher et al.(1999)]{Fisher99} Fisher, A., Branch, D., Hatano, K., \& Baron, E.\ 1999, \mnras, 304, 67
\bibitem[Folatelli et al.(2006)]{Folatelli06} Folatelli, G., et al.\ 2006, \apj, 641, 1039
\bibitem[Galama et al.(1998)]{Galama98} Galama, T.~J., et al.\ 1998, \nat, 395, 670
\bibitem[Gezari et al.(2008)]{Gezari08} Gezari, S., et al.\ 2008, \apj, 683, L131
\bibitem[Ghirlanda et al.(2007)]{Ghirlanda07} Ghirlanda, G., Nava, L., Ghisellini, G., \& Firmani, C. 2007, \aap, 466, 127
\bibitem[Gorosabel et al.(2008)]{Gorosabel08} Gorosabel, J., et al.\ 2008, \apj L, arXiv:0810.4333, submitted
\bibitem[Hjorth et al.(2003)]{Hjorth03} Hjorth, J., et al.\ 2003, \nat, 423, 847
\bibitem[Hirata et al.(1987)]{Hirata87} Hirata, K., et al.\ 1987, \prl, 58, 1490
\bibitem[Li(2007)]{Li07} Li, L.-X.\ 2007, \mnras, 375, 240
\bibitem[Li(2008)]{Li08} Li, L.-X.\ 2008, \mnras, 388, 603
\bibitem[Malesani et al.(2008)]{Malesani08a} Malesani, D., et al.\ 2008, GCN Circ. 7169
\bibitem[Mazzali et al.(2006)]{Mazzali06} Mazzali, P.~A., et al.\ 2006, \nat, 442, 1011
\bibitem[Mazzali et al.(2008)]{Mazzali08} Mazzali, P.~A., et al.\ 2008, Science, 321, 1185
\bibitem[Menzies et al.(1987)]{Menzies87} Menzies, J.~W., et al.\ 1987, \mnras, 227, 39
\bibitem[Modjaz et al.(2008a)]{Modjaz08a} Modjaz, M., Li, W., Butler, N., et al.\ 2008a, \apj, submitted (arXiv:0805.2201)
\bibitem[Modjaz et al.(2008b)]{Modjaz08b} Modjaz, M., Chornock, R., Foley, R.~J., Filippenko, A.~V., Li, W., \& Stringfellow, G.\ 2008b, CBET 1221
\bibitem[Munari \& Zwitter(1997)]{Munari97} Munari, U., \& Zwitter, T.\ 1997, \aap, 318, 269
\bibitem[Patat et al.(2001)]{Patat01} Patat, F., et al.\ 2001, \apj, 555, 900
\bibitem[Predehl \& Schmitt(1995)]{Predehl95} Predehl, P., \& Schmitt, J.~H.~M.~M.\ 1995, \aap, 293, 889
\bibitem[Quimby et al.(2006)]{Quimby06} Quimby, R.~M., Wheeler, J.~C., H\"oflich, P., Akerlof, C.~W., Brown, P.~J., \& Rykoff, E.~S. 2006, \apj, 666, 1093
\bibitem[Quimby et al.(2007)]{Quimby07} Quimby, R.~M., Aldering, G., Wheeler, J.~C., H\"oflich, P., Akerlof, C.~W., \& Rykoff, E.~S. 2007, \apj, 668, L99
\bibitem[Richmond et al.(1994)]{Richmond94} Richmond, M.~W., Treffers, R.~R., Filippenko, A.~V., Paik, Y., Leibundgut, B., Schulman, E., \& Cox, C. V. 1994, \aj, 107, 1022
\bibitem[Richmond et al.(1996)]{Richmond96} Richmond, M.~W., et al.\ 1996, \aj, 111, 327
\bibitem[Sauer et al.(2006)]{Sauer06} Sauer, D.~N., Mazzali, P.~A., Deng, J., Valenti, S., Nomoto, K, \& Filippenko, A.~V. 2006, \mnras, 369, 1939
\bibitem[Shawinski et al.(2008)]{Schawinski08} Schawinski, K., et al.\ 2008, Science, 321, 223
\bibitem[Soderberg et al.(2008a)]{Soderberg08a} Soderberg, A.~M., et al.\ 2008a, \nat, 453, 469
\bibitem[Soderberg et al.(2008b)]{Soderberg08b} Soderberg, A.~M., Berger, E., Fox, D., Cucchiara, A., Rau, A., Ofek, E., Kasliwal, M., \& Cenko, S.~B. 2008b, GCN Circ. 7165
\bibitem[Stanek et al.(2003)]{Stanek03} Stanek, K.~Z., et al.\ 2003, \apj, 591, L17
\bibitem[Stritzinger et al.(2002)]{Stritzinger02} Stritzinger, M., et al.\ 2002, \aj, 124, 2100
\bibitem[Tanaka et al.(2008)]{Tanaka08} Tanaka, M., et al.\ 2008, \apj, arXiv:0807.1674, in press
\bibitem[Th\"one et al.(2008)]{Thoene08} Th\"one, C.~C., Michalowski, M.~J, Leloudas, G., Cox, N.~L.~J., Fynbo, J.~P.~U., Sollerman, J., Hjorth, J., \& Vreeswijk, P.~M. 2008, \apj, arXiv:0807.0473, submitted
\bibitem[Turatto et al.(2003)]{Turatto03} Turatto, M., Benetti, S., \& Cappellaro, E.\ 2003, From Twilight to Highlight: The Physics of Supernovae, ed.\ W. Hillebrandt \& B. Leibundgut (Berlin: Springer), 200
\bibitem[Valenti et al.(2008a)]{Valenti08} Valenti, S., Turatto, M., Navasardyan, H., Benetti, S., \& Cappellaro, S.\ 2008a, GCN Circ. 7163
\bibitem[Valenti et al.(2008b)]{Valenti08b} Valenti, S., D'Elia, V., Della Valle, M., Benetti, S., Chincarini, G., Mazzali, P.~A., \& Antonelli, L.~A. 2008b, GCN Circ. 7221
\bibitem[Waxman et al.(2007)]{Waxman07} Waxman, E., M\'esz\'aros, P., \& Campana, S.\ 2007, \apj, 667, 351
\bibitem[Xu et al.(2008)]{Xu08} Xu, D., Zou, Y.~C., \& Fan, Y.~Z. 2008, \apj, arXiv:0801.4325, submitted

\end{thebibliography}
\end{document}